\documentclass[12pt,preprint]{emulateapj}

\begin{document}

\title{The Stellar Populations of Praesepe and Coma Berenices}

\author{Adam L. Kraus, Lynne A. Hillenbrand}

\affil{California Institute of Technology, Department of Astrophysics, MC 105-24, Pasadena, CA 91125}

\email{(alk@astro.caltech.edu)}
\email{(lah@astro.caltech.edu)}

\begin{abstract}

We present the results of a stellar membership survey of the nearby open 
clusters Praesepe and Coma Berenices. We have combined archival survey data 
from the SDSS, 2MASS, USNOB1.0, and UCAC-2.0 surveys to compile proper 
motions and photometry for $\sim$5 million sources over 300 deg$^2$. Of 
these sources, 1010 stars in Praesepe and 98 stars in Coma Ber are 
identified as candidate members with probability $>$80\%; 442 and 61 are 
identified as high-probability candidates for the first time. We estimate 
that this survey is $>$90\% complete across a wide range of spectral types 
(F0 to M5 in Praesepe, F5 to M6 in Coma Ber). We have also investigated the 
stellar mass dependence of each cluster's mass and radius in order to 
quantify the role of mass segregation and tidal stripping in shaping the 
present-day mass function and spatial distribution of stars. Praesepe shows 
clear evidence of mass segregation across the full stellar mass range; Coma 
Ber does not show any clear trend, but low number statistics would mask a 
trend of the same magnitude as in Praesepe. The mass function for Praesepe 
($\tau$$\sim$600 Myr; $M\sim$500 $M_{\sun}$) follows a power law consistent 
with that of the field present-day mass function, suggesting that any 
mass-dependent tidal stripping could have removed only the lowest-mass 
members ($<$0.15 $M_{\sun}$). Coma Ber, which is younger but much less 
massive ($\tau$$\sim$400 Myr; $M\sim$100 $M_{\sun}$), follows a 
significantly shallower power law. This suggests that some tidal stripping 
has occurred, but the low-mass stellar population has not been strongly 
depleted down to the survey completeness limit ($\sim$0.12 $M_{\sun}$).

\end{abstract}


\section{Introduction}

Star clusters are among the most powerful and versatile tools available to 
stellar astronomy. Nearby clusters serve as prototypical populations for 
studying many diverse topics of stellar astrophysics, including star 
formation, stellar structure, stellar multiplicity, and circumstellar 
processes like planet formation (e.g. Patience et al. 2002; Bouy et al. 
2006; Muench et al. 2007; Stauffer et al. 2007; Siegler et al. 2007); star 
clusters are uniquely sensitive to the physics of these processes due to 
their uniform and well-constrained age, distance, and metallicity. Open 
clusters are also thought to be the birthplaces of most stars, so the 
formation, evolution, and disruption of clusters establish the environment 
of star formation and early stellar evolution.  Two of the nearest open 
clusters are Praesepe and Coma Berenices. Praesepe is a rich ($N\sim1000$ 
known or suspected members), intermediate age ($\sim$600 Myr) cluster at a 
distance of 170 pc (Hambly et al. 1995a), while Coma Ber is younger and 
closer ($\sim$400 Myr; 90 pc) and much sparser ($N\sim150$; Casewell et 
al. 2006).

Praesepe has been the target of numerous photometric and astrometric 
membership surveys over the past century; part of the reason for its 
popularity is that its proper motion is relatively distinct from that of 
field stars (-36.5,-13.5 mas yr$^{-1}$), simplifying the identification of 
new members. Its high-mass stellar population was identified early in the 
last century by Klein-Wassink (1927), and subsequent surveys extended the 
cluster census to intermediate-mass stars (Artyukhina 1966; Jones \& 
Cudworth 1983). The M dwarf stellar population was first identified by 
Jones \& Stauffer (1991). A later survey by Hambly et al. (1995a) extended 
this work to a fainter limit and a larger fraction of the cluster, 
producing a cluster census that is still used for most applications (e.g. 
Allen \& Strom 1995; Holland et al. 2000; Kafka \& Honeycutt 2006). There 
have been additional surveys to identify cluster members, but they have 
been prone to contamination from field stars (Adams et al. 2002) or based 
purely on photometry with no astrometric component (Pinfield et al. 1997; 
Chappelle et al. 2005).

Coma Ber, in contrast, has been largely neglected in surveys of nearby 
open clusters. The cluster would be an ideal population for many studies 
due to its proximity (second only to the Hyades) and intermediate age 
between the Pleiades (125 Myr) and Hyades or Praesepe ($\sim$600 Myr), but 
its members are difficult to distinguish from field stars because it has a 
proper motion (-11.5, -9.5 mas yr$^{-1}$) which is significantly lower 
than that of Praesepe. It is also a much sparser cluster than Praesepe, 
and its few members are projected over a much larger area of the sky. Its 
high-mass stellar population has been known for many decades (Trumpler 
1937), but only a handful of additional members have been confirmed 
(Artyunkhin 1966; Argue \& Kenworthy 1969; Bounatiro 1993; Odenkirchen et 
al. 1998); many candidate members have been identified, but a large 
fraction of them have been shown to be unrelated field stars (e.g. 
Jeffries 1999; Ford et al. 2001). One survey for low-mass stars was 
conducted recently by Casewell et al. (2006), who used 2MASS photometry 
and USNO-B1.0 astrometry to identify 60 candidate members extending well 
into the M dwarf regime ($\sim$0.30 $M_{\sun}$). This survey discovered 
many candidate members with spectral types of late G and early M, but as 
we will discuss later, significant contamination from field stars rendered 
it completely insensitive to $K$ dwarf members and diluted its other 
discoveries with a significant number of nonmembers.

In this paper, we combine the photometric and astrometric results of 
several wide-field imaging surveys to compile a full stellar census of 
Praesepe and Coma Ber. This census is both wider and deeper than any 
previous proper motion survey, extending to near the substellar boundary. 
Our results for Praesepe allow us to fully characterize the structure and 
dynamical evolution of this prototypical cluster, while our results for 
Coma Ber unveil a new benchmark stellar population that is closer than any 
cluster except the Hyades and that fills a poorly-studied age range. In 
Section 2, we describe the all-sky surveys that contribute to our cluster 
census, and in Section 3, we describe the photometric and astrometric 
analysis techniques that we used to identify new members. We summarize our 
new catalog of cluster members in Section 4. Finally, in Section 5, we 
analyze the structure and properties of each cluster.

\section{Data Sources}

In this survey, we worked with archival data from several 
publicly-available surveys: SDSS, 2MASS, USNO-B1.0, and UCAC2. In each 
case, we extracted a portion of the source catalogue from the data access 
websites. We worked with circular areas of radius 7$^o$ centered on the 
core of each cluster (8h40m, +20$^o$ and 11h24m,+26$^o$, respectively); 
for both clusters, this radius is approximately twice the estimated tidal 
radius (Hambly et al. 1995a; Casewell et al. 2006).

\subsection{SDSS}

The Sloan Digital Sky Survey (SDSS; York et al. 2000) is an ongoing deep 
optical imaging and spectroscopic survey of the northern galactic cap. The 
most recent data release (DR5; Adelman-McCarthy et al. 2007) reported 
imaging results in five filters ($ugriz$) for 8000 deg$^2$, including the 
full areas of Praesepe and Coma Ber. The 10$\sigma$ detection limits in 
each filter are $u=22.0$, $g=22.2$, $r=22.2$, $i=21.3$, and $z=20.5$; the 
saturation limit in all filters is $m\sim14$. The typical absolute 
astrometric accuracy is $\sim$45 mas rms for sources brighter than $r=20$, 
declining to 100 mas at $r=22$ (Pier et al. 2003); absolute astrometry was 
calibrated with respect to stars from UCAC2, which is calibrated to the 
Inertial Coordinate Reference Frame (ICRS).

The default astrometry reported by the SDSS catalog is the $r$ band 
measurement, not the average of all five filters. However, the residuals 
for each filter (with respect to the default value) are available, so we 
used these residuals to construct a weighted mean value for our analysis. 
We adopted a conservative saturation limit of $m\sim$15 in all filters, 
even though the nominal saturation limit is $m\sim$14, because we found 
that many photometric measurements were mildly saturated for $14<m<14.5$. 
We also neglect measurements which are flagged by the SDSS database as 
having one or more saturated pixels. Finally, we removed all sources which 
did not have at least one measurement above the nominal 10$\sigma$ 
detection limits. Any cluster members fainter than this limit will not 
have counterparts in other catalogs, and the presence of excess sources 
can complicate attempts to match counterparts between datasets.

\subsection{USNO-B1.0}

The USNO-B1.0 survey (USNOB; Monet et al. 2003) is a catalogue based on 
the digitization of photographic survey plates from five epochs. For 
fields in the north, including both Praesepe and Coma Ber, these plates 
are drawn from the two Palomar Observatory Sky Surveys, which observed the 
entire northern sky in the 1950s with photographic B and R plates and the 
1990s with photographic B, R, and I plates; we follow standard USNOB 
nomenclature in designating these observations $B1$, $R1$, $B2$, $R2$, and 
$I2$. 

The approximate detection limits of the USNOB catalog are $B\sim$20, 
$R\sim$20, and $I\sim$19, and the observations saturate for stars brighter 
than $V\sim$11. The typical astrometric accuracy at each epoch is 
$\sim$120 mas, albeit with a significant systematic uncertainty (up to 200 
mas) due to its uncertain calibration into the the ICRS via the 
unpublished USNO YS4.0 catalog. As we describe in Section 3.2, we have 
recalibrated the USNOB astrometry at each epoch using UCAC2 astrometry; 
this step reduces the systematic uncertainty.

\subsection{2MASS}

The Two-Micron All-Sky Survey (2MASS; Skrutskie et al. 2006) observed the 
entire sky in the $J$, $H$, and $K_s$ bands over the interval of 
1998-2002. Each point on the sky was imaged six times and the coadded 
total integration time was 7.8s, yielding 10$\sigma$ detection limits of 
$K=14.3$, $H=15.1$, and $J=15.8$. The saturation levels depend on the 
seeing and sky background for each image, but are typically $J<9$, 
$H<8.5$, and $K_{s}<8$. However, the NIR photometry is typically accurate 
to well above these saturation limits since it was extrapolated from the 
unsaturated PSF wings. The typical astrometric accuracy attained for the 
brightest unsaturated sources ($K\sim8$) is $\sim$70 mas. The absolute 
astrometry calibration was calculated with respect to stars from Tycho-2; 
subsequent tests have shown that systematic errors are typically $\la$30 
mas (Zacharias et al. 2003).

\subsection{UCAC2}

The astrometric quality of all three of the above surveys could be 
compromised for bright, saturated stars, so proper motions calculated from 
those observations could be unreliable. Many of the brightest stars are 
saturated in all epochs, so we have no astrometry with which to compute 
proper motions. We have addressed this problem by adopting proper motions 
for bright stars as measured by the Second USNO CCD Astrograph Catalog 
(UCAC2; Zacharias et al. 2004).

UCAC2 was compiled from a large number of photographic sky surveys and a 
complete re-imaging of the sky south of $\delta\sim40^o$. UCAC2 is not 
complete since many resolved sources (double stars and galaxies) were 
rejected. However, most sources between $R=8$ and $R=16$ should be 
included. The typical errors in the reported proper motions are $\sim$1-3 
mas yr$^{-1}$ down to $R=12$ and $\sim$6 mas yr$^{-1}$ to $R=16$. We have 
adopted UCAC2 proper motions in cases where we were unable to calculate 
new values or where the UCAC2 uncertainties are lower than the 
uncertainties for our values.

\subsection{Known Members of Praesepe}

There have been many previous surveys to identify members of Praesepe, so 
we have compiled a list of high-confidence cluster members that can be 
used to test our survey procedures (Section 3) and determine the 
completeness of our survey (Section 4.2). We have not done the same for 
Coma Ber since there are far fewer high-confidence members ($<$50). 
However, the brightness ranges are similar enough that the detection 
efficiencies should be similar for both clusters.

We drew our high-confidence Praesepe sample from the proper motion surveys 
of Jones \& Cudworth (1983), Jones \& Stauffer (1991), and Hambly et al. 
(1995a). We also included the high-mass stars identified by Klein-Wassink 
(1927) which possessed updated astrometry in the survey by Wang et al. 
(1995). We required each member of our high-confidence sample to have been 
identified with $\ge$95\% probability of membership by at least one 
survey, and to not have been identified with $<$80\% probability by any 
other survey; a total of 381 sources met these requirements.

\section{Data Analysis}

Cluster surveys typically identify candidate members using a combination 
of photometric and astrometric data. All cluster members have the same 
age, distance, and 3-D spatial velocity, so they follow the same 
color-magnitude sequence and have the same proper motion. This allows for 
the efficient rejection of all nonmembers which do not meet both criteria.

In the following subsections, we describe our procedure for applying these 
tests. First, we use SED fitting for our photometric data (spanning 
0.3-2.3 $\mu$m) to estimate the temperatures and luminosities of all 
$\sim$5 million sources, and then we calculate a weighted least-squares 
fit of our time-series astrometric data to calculate the corresponding 
proper motions. After deriving both sets of results, we then cut the 
overwhelming majority of sources which do not follow the cluster 
photometric sequence. Finally, we examine the (much smaller) list of 
remaining sources and determine membership probabilities based on the 
level of agreement between individual candidate astrometry (proper motion 
and radius from cluster center) and the corresponding distributions for 
the cluster and for background stars.

We chose to apply the cuts in this order specifically because the final 
membership probabilities are based on the astrometric properties and not 
the photometric properties, but inverting the order of the cuts would not 
affect our final results. Both sets of tests were crucial in narrowing the 
list of candidates. Of the $\sim$10$^6$ sources in each cluster for which 
we measured proper motions, $\sim10^5$ would have been selected by a 
purely kinematic test and $\sim10^4$ would have been selected by a purely 
photometric test.

\subsection{SED Fitting}

\begin{figure*}
\epsscale{1.00}
\plotone{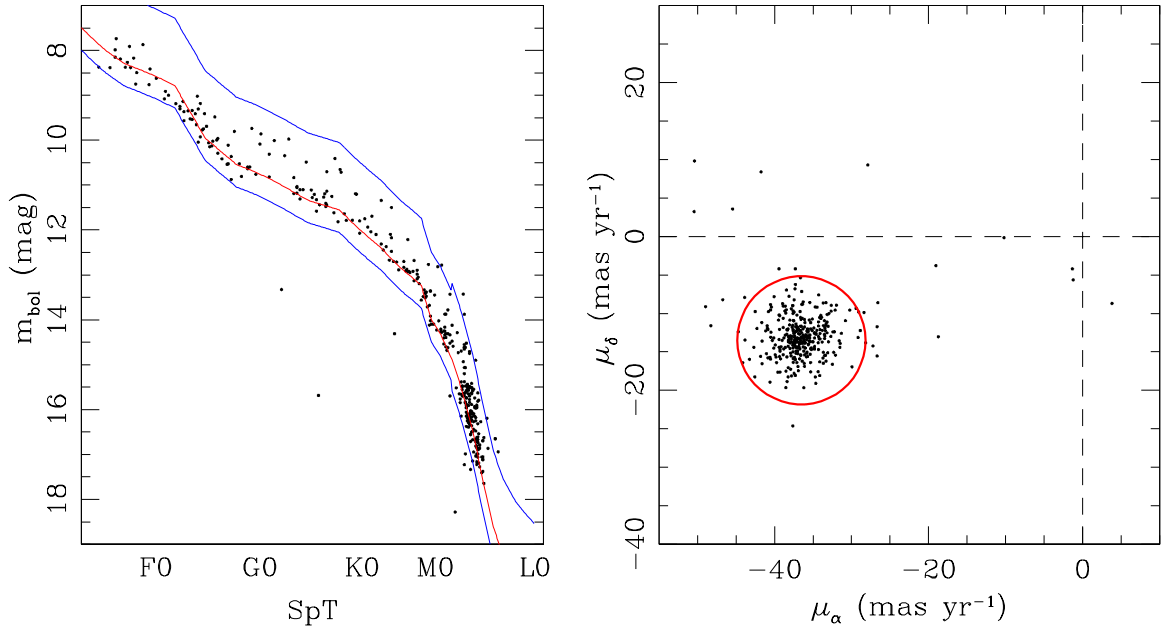}
\caption{HR and proper motion diagrams for our high-confidence sample of 
Praesepe members. For the HR diagram, we plot the cluster single-star 
sequence (red) and the selection range for identifying new members 
(blue). In the proper motion diagram, we plot a circle of radius 8 
mas yr$^{-1}$ (approximately 2$\sigma$ for a typical M4 member) centered 
at the mean cluster proper motion.}
\end{figure*}

\begin{figure}
\epsscale{1.00}
\plotone{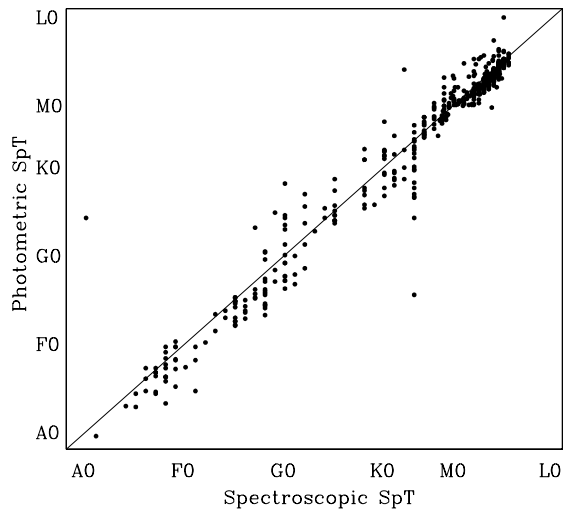}
\caption{A comparison of our photometric spectral type determinations to 
spectroscopic determinations for 632 candidate Praesepe members in the 
literature. The small excess of points below the relation at spectral 
type K3 are all drawn from the spectroscopic survey of Adams et al. 
(2002), which observed spectra in a red wavelength range that contained 
no diagnostics for distinguishing FGK stars. The A0 star that we 
misclassified (KW 552) is an Algol-type eclipsing binary, so the 2MASS 
photometry may have been obtained during primary eclipse; we did not use 
any SDSS photometry in its SED fit because it was all saturated. If this is 
the case, our derived spectral type corresponds to an unknown combination 
of light from the primary and secondary. The K2 star that we misclassified 
(KW 572) was biased by saturated SDSS photometry which was not flagged.}
\end{figure}

We base our photometric analysis on the merged results from 2MASS and 
SDSS, which yield measurements in 8 filters ($ugrizJHK$) for each source. 
We do not use the photometric results reported by USNOB because they are 
much more uncertain ($\sim$0.25 mag) and do not introduce any new 
information beyond that reported by SDSS. We also note that many 
high-mass sources were saturated in one or more filters, so they had 
fewer than 8 photometric measurements available; the highest-mass stars 
were saturated in all five SDSS filters, leaving only $JHK$ photometry.

Candidate cluster members traditionally have been selected by photometric 
surveys which measure magnitudes in several bandpasses and then estimate 
each star's intrinsic properties (bolometric flux and temperature) using 
its observed properties (magnitudes and colors). Candidate members are 
then selected from those stars which fall along the cluster sequence (as 
defined by known members and by theoretical models) on color-magnitude 
diagrams. However, this method suffers from serious flaws. A single 
magnitude is typically taken as a proxy for flux, which places excessive 
weight on that bandpass and underweights other bandpass(es) in the survey. 
If there are more than two bandpasses, motivating the use of multiple 
CMDs, then color-magnitude selection also neglects the covariance between 
measurements, artificially inflating the uncertainty in an object's 
intrinsic properties. Finally, the use of many CMDs introduces significant 
complexity in the interpretation and communication of results.

We have addressed these challenges by developing a new method for 
photometric selection of candidate members. Instead of using many 
different combinations of color and magnitude as proxies for stellar flux 
and temperature, we have used an SED fitting routine to estimate directly
each star's intrinsic properties, then selected candidate members based on 
their positions in the resulting HR diagram. This method is not vulnerable 
to the flaws of individual color-magnitude selection since it uses all 
data simultaneously and uniformly, and since we can implement it as a 
least-squares minimization, it significantly reduces the uncertainty in 
the final results.

Specifically, for each star we calculated the $\chi$$^2$ goodness of fit 
for the system of eight equations:

\[M_{i}-m_{i}=DM\]

where $m_i$ is the observed magnitude in filter $i$, $M_{i}$ is the 
absolute magnitude in filter $i$ for the SED model being tested, and $DM$ 
is the distance modulus, which was estimated from a weighted least-squares 
fit across all filters. This system ignores the effects of reddening, but 
this should be minimal for both clusters. Taylor et al. (2006) found a 
reddening value for Praesepe of $E(B-V)=27\pm4$ mmag, while Feltz (1972) 
found a value for the Coma Ber region of $E(B-V)=0\pm2$ mmag.

We tested a library of 491 stellar SEDs which spanned a wide range of 
spectral types: B8 to L0, in steps of 0.1 subclasses. We describe the SED 
library and its construction in more detail in Appendix A. We rejected 
potentially erroneous observations by rejecting any measurement that 
disagreed with the best-fit SED by more than 3$\sigma$, where $\sigma$ is 
the photometric error reported by the SDSS or 2MASS, and then calculating 
a new fit. The model which produced the best $\chi$$^2$ fit over the 8 
filters was adopted as the object spectral type, and the corresponding 
value of $DM$ was added to the model's absolute bolometric magnitude to 
estimate the apparent bolometric flux. The uncertainties in the spectral 
type and distance modulus were estimated from the 1$\sigma$ interval of 
the $\chi$$^2$ fit for each object.

In the left panel of Figure 1, we plot an H-R diagram for our 
high-confidence sample of Praesepe members. The red line shows the field 
main sequence at the distance of Praesepe (Appendix A), and the blue lines 
show the upper and lower limits that we use for identifying cluster 
members. For stars earlier than M2, these limits are set 0.5 magnitudes 
below and 1.5 magnitudes above the main sequence to allow for the width of 
the cluster sequence (due to errors, the finite depth of the cluster, and 
the existence of a multiple-star sequence). The late main sequence is 
nearly vertical in the HR diagram, which suggests that uncertainties in 
spectral type will be more important than uncertainties in flux for 
broadening the cluster sequence. We account for this by extending the 
selection range for spectral types $\ge$M2 to 0.7 magnitudes below and 1.7 
magnitudes above the field main sequence. Most of the 15 outliers have 
fluxes or spectral types that are biased by one or more photometric 
measurements which appear to be erroneous by less than $3\sigma$, causing 
them to fall just outside our selection range. However, four sources 
appear to have colors and magnitudes that are genuinely inconsistent with 
the cluster sequence.

In Figure 2, we plot our photometric spectral type against 
previously-measured spectroscopic spectral types for 632 candidate 
Praesepe members (Ramberg 1938; Bidelman 1956; Corbally \& Garrison 1986; 
Abt 1986; Williams et al. 1994; Allen \& Strom 1995; Adams et al. 2002; 
Kafka \& Honeycutt 2006). The two sets of spectral types agree 
systematically to within $<$2 subclasses; the dispersion in the relation 
is $\sim$3 subclasses for early-type stars (A0-G0) and $\la$1 subclass for 
later-type stars (G0-M6). This dispersion represents the combined 
dispersions of both our measurements and those in the literature, so it 
represents an upper limit on the statistical uncertainties in our spectral 
type estimate. Most of the early-type stars were classified by Ramberg and 
Bidelman, so the larger scatter could be a result of their older, less 
precise observing techniques. However, our SED-fitting routine rejected 
most of the SDSS photometry for these sources since it was saturated, so 
some of the uncertainty may be a result of using only 2MASS $JHK$ 
photometry.

When applied to our full source list, our photometric selection criteria 
identify 11,999 candidate members of Praesepe and 2,034 candidate members 
of Coma Ber. As we demonstrate in the Section 3.2 and 3.3, the vast 
majority of these sources are probably background stars since they have 
proper motions inconsistent with cluster membership.

\subsection{Proper Motions}

\begin{deluxetable}{lrr}
\tabletypesize{\scriptsize}
\tablewidth{0pt}
\tablecaption{Astrometric Recalibration Offsets}
\tablehead{\colhead{Cluster/Epoch} & \colhead{$\Delta_{\alpha}$}
& \colhead{$\Delta_{\delta}$}
}
\startdata
Praesepe B1&+42&+97\\
Praesepe R1&+49&+104\\
Praesepe B2&+10&-75\\
Praesepe R2&-2&-78\\
Praesepe I2&-11&-119\\
Coma Ber B1&-16&+55\\
Coma Ber R1&-21&+80\\
Coma Ber B2&-132&-58\\
Coma Ber R2&-96&-64\\
Coma Ber I2&-118&-93\\
\enddata
\tablecomments{Offsets are measured in mas. The typical uncertainty for 
each offset, as estimated from the standard deviation of the mean, is 
$\sim$3-5 mas.}
\end{deluxetable}

Kinematic measurements are a key tool in identifying members of stellar 
populations. Internal cluster velocity dispersions are typically much 
lower than the dispersion of field star velocities, so stellar populations 
generally can be distinguished from the field star population by their 
uniform kinematics. The measurement of tangential kinematics, via proper 
motions, is also an efficient method since it can be applied to many 
cluster members simultaneously using wide-field imaging. Many recent 
efforts have employed various combinations of all-sky surveys in order to 
systematically measure proper motions of both clusters and field stars; 
USNOB is itself a product of such analysis, and Gould \& Kohlmeier (2004) 
produced an astrometric catalog for the overlap between USNOB and SDSS 
Data Release 1. However, there has been no systematic attempt to combine 
all available catalogs using a single algorithm to produce a single 
unified set of kinematic measurements.

Before calculating proper motions for our survey, our first step was to 
recalibrate the five epochs of USNOB astrometry into the ICRS. The densest 
reference system that is directly tied to the ICRS is UCAC2, which we 
already cross-referenced with our dataset, so we used all of its sources 
with high-precision astrometry ($\sigma_{\mu}$$\la$4 mas yr$^{-1}$) as 
calibrators. For each USNOB epoch, we projected the simultaneous UCAC2 
positions of all calibrators using modern (epoch 2000) UCAC2 astrometry 
and proper motions, then determined the median offset between the 
predicted UCAC2 values and the observed USNOB values. These offsets were 
then added to each USNOB source to bring its astrometry into the ICRS. We 
list these mean offsets in Table 1; each offset was typically calculated 
from $\sim$3000 sources, and the standard deviation of the mean for each 
offset was $\sim$3-5 mas. The median offsets were small ($<$150 mas), so 
the net change in our calculated final proper motions is $\la$3 mas 
yr$^{-1}$.

After we recalibrated all surveys into the same reference system, we used 
a weighted least-squares fit routine to calculate the proper motion of 
each object based on all available astrometry for unsaturated detections. 
Our algorithm tested the goodness of each fit and rejected all outliers at 
$>3\sigma$; most of these outliers were found in the photographic survey 
data, not in 2MASS or SDSS.

In the right panel of Figure 1, we plot a proper motion diagram for our 
high-confidence sample of Praesepe members. The mean cluster proper motion 
(-36.5,-13.5 mas yr$^{-1}$) is denoted by a red circle with a radius of 8 
mas yr$^{-1}$ (twice the typical 1$\sigma$ uncertainty for the M4 members 
in our high-confidence sample). We found that 326 of our 381 
high-confidence members fall within this limit, and most of the early-type 
stars (which have much smaller errors) form a much tighter distribution. 
Most of the outliers appear to be biased by erroneous first-epoch 
positions that can not be rejected at a 3$\sigma$ level by our fitting 
routine. These early epochs are not significantly more prone to erroneous 
measurements than later photographic measurements, but they change the 
resulting proper motion by a larger amount since their time baseline with 
respect to all other measurements is so long.

Our subsequent kinematic analysis (Section 3.3) has retained all 
photometric candidates with proper motions within 20 mas yr$^{-1}$ 
(5$\sigma$ for low-mass candidates) of each cluster's mean proper motion; 
we set this limit to be much larger than the cluster distribution so that 
we would also retain enough field stars to determine their density in 
proper motion space. We found that 2611 of our 11999 photometric candidates 
in Praesepe and 645 of our 2034 photometric candidates in Coma Ber fell 
within this limit. 

We removed a small number of sources (44 from Praesepe and 4 from Coma Ber) 
that had highly uncertain proper motions ($\sigma$$>$10 mas yr$^{-1}$) 
because we could not have accurately assessed their membership. The 
astrometry was typically more uncertain for these few sources because there 
were few or no detections in USNOB. We also visually inspected the SED for 
any source with a poor photometric fit ($\chi^2_{\nu}>10$) and rejected two 
sources near Coma Ber which were only selected due to saturated SDSS 
photometry that had not been flagged.

Finally, we visually inspected the color-composite SDSS image of each 
source using the SDSS batch image 
service\footnote{http://cas.sdss.org/dr5/}. We found that 8 sources in 
Praesepe and 31 sources in Coma Ber were resolved background galaxies, so 
we removed them from further consideration. These galaxies were split 
roughly evenly between bright ($r\sim14-16$) sources with K star colors and 
faint ($r\sim19$) galaxies with red $riz$ colors and no $ug$ or $JHK$ 
detections; in all cases, the apparent proper motion was caused by a large 
scatter in the photometric centroids. The SDSS database also includes 
a morphological classification of whether each object is a star or galaxy 
that is likely to be more sensitive than visual inspection, but we have 

found that saturated stars and marginally resolved binaries are often 
classified as galaxies by the SDSS pipeline, so we chose not to use this 
parameter in rejecting likely galaxies. 

\subsection{Identification of Cluster Members}

Our photometric and astrometric selection criteria do not perfectly reject 
field stars, so we expect that some fraction of our candidates will 
actually be interlopers and not cluster members. Many surveys quantify the 
level of contamination by studying one or more control populations, 
selected from a nearby volume of kinematic or spatial parameter space. The 
membership probability for a set of stars is then represented by the 
fractional excess in the candidate population with respect to the control 
population. However, this choice ignores all information about the spatial 
or proper motion distribution of the candidates, treating these 
distributions as constant within the selection limits. A more rigorous 
approach should take these non-constant probability density functions into 
account, giving highest membership probability to those candidates that 
are closest to the cluster center and have proper motions closest to the 
mean cluster value.

To this end, we have adopted the maximum likelihood method of Sanders 
(1971) and Francic (1989) to distinguish cluster members and field stars 
among the candidates that meet our photometric and kinematic selection 
criteria. This method explicitly fits the spatial and kinematic 
distributions of all candidates with two separate probability density 
functions, $\Phi=\Phi_{c}+\Phi_{f}$, corresponding to cluster members and 
field interlopers. The method then assigns a membership probability to 
each star based on the values of each distribution for that location in 
parameter space, $P_{mem}=\Phi_{c}/(\Phi_{c}+\Phi_{f})$.

Following some of the refinements of Francic (1989), we chose to fit the 
cluster spatial distribution with an exponential function and the cluster 
proper motion distribution with a gaussian function:

\[\Phi_{c}(\mu_{\alpha},\mu_{\delta},r)=
\frac{N_{c}e^{-r/r_{0}}}{2\pi^2r_0^2\sigma^2} 
e^{\frac{1}{2\sigma^2}((\mu_{\alpha}-\mu_{\alpha,m})^2+
(\mu_{\delta}-\mu_{\delta,m})^2)}\]

Where the quantities $N_c$ (the total number of cluster stars), 
$r_{0}$ (the scale radius), and $\sigma$ (the standard deviation of 
the cluster proper motion distribution) were determined from the fit. 
We adopted the mean proper motions of each cluster, 
$(\mu_{\alpha,m},\mu_{\delta,m})=(-36.5,-13.5)$ mas yr$^{-1}$ 
(Praesepe) and $(-11.5,-9.5)$ mas yr$^{-1}$ (Coma Ber), from the 
literature; these results match UCAC2 values for known high-mass 
cluster members.

We evaluated the option of fitting the cluster spatial distribution with a 
mass-dependent King profile (King 1962), but we found that the function 
produced a poor fit at large separations. High-mass stars in particular 
are more centrally concentrated than a King profile would predict. By 
contrast, an exponential radial density profile can accurately match the 
outer density profile at the cost of moderately overestimating the central 
density. We decided that it is more important to accurately predict the 
spatial structure of the outer cluster, where cluster members are less 
numerous and harder to distinguish from field stars, so we chose to use 
the exponential profile.

We chose to fit the field spatial distribution with a constant function 
since the density of field stars does not vary significantly at these high 
galactic latitudes. In a departure from previous convention, we also chose 
to fit the field proper motion distribution with a constant function. As 
we show in Figures 4 and 5, the proper motion distribution of field stars 
is not easily parametrized with a single function. However, the 
distribution varies only on scales much larger than the astrometric 
precision for typical mid-M candidates ($\sim$4 mas yr$^{-1}$). If we 
consider a small region of parameter space, then the distribution should 
be roughly constant. Thus, the field probability density function we have 
adopted is:

\[\Phi_{f}= \frac{N_{total}-N_{c}}{A_{SP}A_{PM}}\]

Where $N_{total}$ is the total number of stars (field and cluster), $N_c$ 
is the number of cluster stars, $A_{SP}$ represents the total spatial area 
of our survey on the sky (a circle with radius 7$^o$), and $A_{PM}$ 
represents the total area of proper motion parameter space from which we 
selected candidates (a circle with radius 20 mas yr$^{-1}$). The proper 
motion criterion was chosen to be much larger than the typical uncertainty 
in cluster proper motions ($\sim$5$\sigma$ for the faintest stars) while 
being small enough that an assumption of a constant field distribution is 
approximately valid. 
 
Both clusters are old enough for mass segregation to have occurred, plus 
the astrometric uncertainties depend significantly on brightness, so we 
expect that the spatial and kinematic distributions will show a 
significant mass dependence. We have accounted for this by dividing each 
cluster sample into spectral type bins and fitting these bins 
independently. As we describe in Section 5, this choice also offers a 
natural system for quantifying the mass-dependent properties of each 
cluster. Our parametrization of the cluster spatial and proper motion 
distributions provides direct measurements of the cluster mass function 
(via $N_c$), the astrometric precision (via $\sigma$), and the effects of 
mass segregation (via $r_0$).

Finally, we determined confidence intervals for each value via a bootstrap 
Monte Carlo routine. This method creates synthetic datasets by drawing 
with replacement from the original dataset; for each bin we constructed 
100 synthetic datasets with the same number of total members, re-ran our 
analysis for each set, and used the distribution of results to estimate 
the standard deviations of the fit parameters.

In Table 2, we summarize the parameter fits. We found in both clusters 
that the fits for spectral types $>$M6 predicted marginally significant 
values of $N_c$, a result we attribute to our nondetection of most 
late-type members. We therefore will not use those parameters in our 
analysis of the mass-dependent cluster properties. However, in the 
interest of completeness, we will still report any candidates which have 
high membership probabilities. Some of these stars have already been 
identified as candidates by previous surveys (e.g. IZ072; Pinfield et al. 
2003), so they may be worthy of consideration in future studies. We also 
found extremely high contamination rates for K stars in Coma Ber; this is 
a natural result of its low proper motion, which causes confusion with 
background K giants. There are few high-probability K-type members 
identified for Coma Ber, but the fits for bulk properties ($N_c$, $r_0$, 
and $\sigma$) are statistically significant.

\begin{deluxetable}{crrrr}
\tabletypesize{\scriptsize}
\tablewidth{0pt}
\tablecaption{Cluster Fit Parameters}
\tablehead{\colhead{SpT} & \colhead{$N_c$} & \colhead{$N_{tot}$} & 
\colhead{$r_0$ (deg)} & \colhead{$\sigma$ (mas yr$^{-1}$}
}
\startdata
\multicolumn{5}{c}{Praesepe}\\
A-F&89$\pm$9&248&0.45$\pm$0.04&1.36$\pm$0.10\\
G&69$\pm$8&236&0.49$\pm$0.05&1.65$\pm$0.14\\
K0.0-K3.9&72$\pm$9&212&0.66$\pm$0.09&3.44$\pm$0.36\\
K4.0-K7.9&102$\pm$9&247&0.71$\pm$0.06&3.34$\pm$0.16\\
M0.0-M1.9&127$\pm$9&283&0.71$\pm$0.04&2.85$\pm$0.16\\
M2.0-M2.9&90$\pm$10&243&0.92$\pm$0.10&3.03$\pm$0.23\\
M3.0-M3.9&202$\pm$12&440&0.71$\pm$0.03&3.01$\pm$0.17\\
M4.0-M4.9&249$\pm$15&514&0.87$\pm$0.04&4.69$\pm$0.28\\
M5.0-M5.9&40$\pm$6&94&0.80$\pm$0.10&6.30$\pm$0.66\\
M6.0-M6.9&15$\pm$6&42&0.98$\pm$0.38&7.00$\pm$1.93\\
\multicolumn{5}{c}{Coma Ber}\\
A-F&17$\pm$3&25&1.19$\pm$0.24&1.22$\pm$0.19\\
G&13$\pm$3&31&1.06$\pm$0.16&1.19$\pm$0.18\\
K&40$\pm$13&413&1.58$\pm$0.17&3.91$\pm$0.89\\
M0.0-M2.9&24$\pm$5&50&1.33$\pm$0.12&4.58$\pm$0.58\\
M3.0-M5.9&36$\pm$6&78&1.46$\pm$0.12&5.07$\pm$0.58\\
M6.0-M8.9&3$\pm$2&15&1.62$\pm$0.55&4.63$\pm$1.26\\
\enddata
\end{deluxetable}

\section{Results}

\subsection{New Cluster Members}

Based on our kinematic and photometric selection procedures, we 
identified 1130 candidate members of Praesepe and 149 candidate members 
of Coma Ber with membership probabilities of $\ge$50\%; 1010 and 98 of 
these candidates have membership probabilities of $>$80\%. Of these 
high-probability candidates, 76 and 50 are newly-identified as 
proper-motion candidates, while 568 and 37 have been classified as 
high-probability ($>$80\%) candidates in at least one previous survey and 
366 and 11 were previously identified with lower probability (references 
in Section 1). In Tables 3 and 4, we list all candidate members with 
$P_{mem}>$50\%. We also list their derived stellar properties, proper 
motions, membership probabilities, cross-identifications with previous 
surveys, and spectroscopically-determined spectral types. In Figure 3, we 
plot a histogram of the number of candidates as a function of $P_{mem}$ 
for each cluster; a majority of candidates have membership probabilities 
of $>$90\% or $<$10\%, suggesting that most of these candidates are being 
unambiguously identified.

To demonstrate the impact of our selection techniques, in Figure 4 we plot 
an HR diagram for all stars near Praesepe which fall within 2$\sigma$ of 
the mean cluster proper motion (left) and a proper motion diagram for all 
stars which passed our photometric selection criteria (right). In both 
cases, the distribution of cluster members can be visually distinguished 
from the underlying distribution of field stars. However, there is also 
significant overlap between cluster members and field stars, indicating 
that both tests were necessary. The proper motion test was a far better 
discriminant against field stars, a result of Praesepe's high and distinct 
proper motion; the photometric criteria accepted 11,999 sources, but only 
1,932 stars fell within 2$\sigma$ of the cluster's mean proper motion.

Based on the HR diagram, it appears that most field stars with consistent 
proper motions are nearby dwarfs; this is not surprising since few distant 
stars will have the large transverse velocities required to match the 
angular velocity of Praesepe. Based on the proper motion diagram, it 
appears that the interlopers which pass our photometric criteria are split 
evenly between stationary sources (such as halo giants) and moving sources 
with larger, randomly distributed proper motions (disk dwarfs that occupy 
the same physical volume as Praesepe). We also note that a clear binary 
sequence can be seen for early-type stars in the HR diagram, but it blends 
with the sigle-star sequence for late-type stars ($\ga$M0).

In Figure 5, we plot similar HR and proper motion diagrams for the stars 
of Coma Ber. The cluster's HR sequence and proper motion distribution are 
not as visually distinctive since the cluster population is smaller, but 
the combination of kinematics and photometry still allow for the efficient 
identification of candidate members. Unlike for Praesepe, the photometric 
test was a better discriminant (accepting 2,034 sources) than the proper 
motion test (21,264 sources); this is a result of the cluster's lower 
distance (which places it higher in the HR diagram relative to the field 
star population) and much smaller proper motion (which allows more 
contamination from nonmoving background sources).

The HR diagram for Coma Ber (which shows kinematically selected sources) 
includes a recognizeable giant branch and many faint (distant) early-type 
stars, both classes which typically have small proper motions. The proper 
motion diagram, which shows photometrically selected stars, includes far 
fewer sources than Praesepe; again, these are split between nonmoving 
background giants and nearby disk dwarfs. A probable binary sequence can 
also be seen for Coma Ber, though it is not as visually distinctive as for 
Praesepe.

\begin{deluxetable*}{lrrrrrrl}
\tabletypesize{\scriptsize}
\tablewidth{0pt}
\tablecaption{Candidate Members of Praesepe}
\tablehead{\colhead{ID} & \colhead{SpT} & \colhead{$m_{bol}$} &
\colhead{$\mu_{\alpha}$} & \colhead{$\mu_{\delta}$} & 
\colhead{$\sigma_{\mu}$} & \colhead{$P_{mem}$} & \colhead{Previous ID\tablenotemark{a}}
\\
\colhead{} & \colhead{} & \colhead{(mag)} & 
\multicolumn{3}{c}{(mas yr$^{-1}$)} & \colhead{(\%)}
}
\startdata
2MASS J08374071+1931064&A8.0$\pm$3.2&8.17$\pm$0.02&-34.8&-12.5&0.7&99.9&KW 45 (A9; Abt 1986)\\
2MASS J08430594+1926153&F9.5$\pm$3.2&9.74$\pm$0.01&-36.6&-13.8&0.9&99.9&KW495 (F8; Ramberg 1938)\\
2MASS J08393837+1926272&K1.5$\pm$1.0&12.10$\pm$0.01&-33.0&-9.6&1.9&99.2&KW198 (K3; Allen \& Strom 1995)\\
2MASS J08325566+1843582&K3.3$\pm$0.5&12.63$\pm$0.01&-38.1&-12.1&3.0&97.1&JS 17\\
2MASS J08380730+2026557&M1.5$\pm$0.1&14.59$\pm$0.01&-41.4&-13.2&3.0&99.5\\
2MASS J08455917+1915127&M3.5$\pm$0.1&15.56$\pm$0.01&-41.8&-11.0&2.7&96.6&AD 3470 (M4; Adams et al. 2002)\\
2MASS J08410334+1837159&M6.8$\pm$0.2&17.47$\pm$0.01&-37.3&-14.2&4.0&96.5&IZ072 (M4.5; Adams et al. 2002)\\
\enddata
\tablecomments{The full version of Table 3 will be published as an 
online-only table in AJ, and is included at the end of this document.}
\tablenotetext{a}{The survey by Adams et al. (2002) used standard 2MASS names for their sources. We already 
provide these names in the first column, so we have labelled the sources as AD $NNNN$ (where $NNNN$ represents the 
number of the entry in their results table) in the interest of brevity.}
\end{deluxetable*}

\begin{deluxetable*}{lrrrrrrl}
\tabletypesize{\scriptsize}
\tablewidth{0pt}
\tablecaption{Candidate Members of Coma Ber}
\tablehead{\colhead{ID} & \colhead{SpT} & \colhead{$m_{bol}$} &
\colhead{$\mu_{\alpha}$} & \colhead{$\mu_{\delta}$} & 
\colhead{$\sigma_{\mu}$} & \colhead{$P_{mem}$} & \colhead{Previous ID\tablenotemark{a}}
\\
\colhead{} & \colhead{} & \colhead{(mag)} & 
\multicolumn{3}{c}{(mas yr$^{-1}$)} & \colhead{(\%)}
}
\startdata
2MASS J12230841+2551049&F9.7$\pm$2.9&8.97$\pm$0.01&-10.0&-8.5&0.7&100.0&Tr 97 (F8; Abt \& Levato 1977)\\
2MASS J12272068+2319475&G7.9$\pm$1.5&9.91$\pm$0.01&-11.6&-8.8&0.7&99.6&CJD 6 (K0; SIMBAD)\\
2MASS J12262402+2515430&K2.8$\pm$0.5&11.55$\pm$0.02&-15.9&-6.1&1.7&84.9&\\
2MASS J12225942+2458584&K5.4$\pm$0.7&10.86$\pm$0.02&-8.7&-12.3&0.9&89.5&\\
2MASS J12241088+2359362&M2.2$\pm$0.1&14.03$\pm$0.01&-9.9&-9.4&2.7&98.1&CJD 46\\
2MASS J12163730+2653582&M2.6$\pm$0.1&14.04$\pm$0.01&-7.8&-10.9&3.0&97.6&CJD 45\\
\enddata
\tablecomments{The full version of Table 4 will be published as an 
online-only table in AJ, and is included at the end of this document.}
\tablenotetext{a}{The survey by Casewell et al. (2006) did not give explicit names for their sources, so we have 
labelled the sources as CJD $NN$ (where $NN$ represents the number of the entry in their results table).}
\end{deluxetable*}

\begin{figure}
\epsscale{1.00}
\plotone{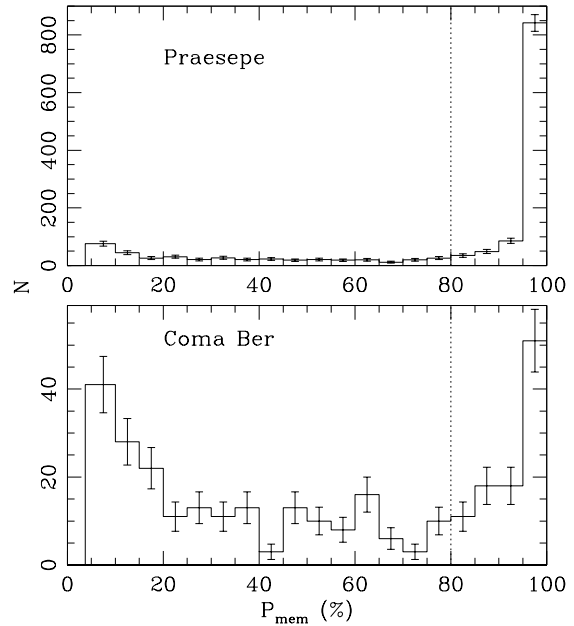}
\caption{The number of candidate members with membership probability 
$P_{mem}$ for Praesepe (top) and Coma Ber (bottom). Most of the Coma Ber 
candidates with $20\%<P_{mem}<80\%$ are K stars, corresponding to the large 
number of candidates which we cannot conclusively distinguish as either K 
dwarf members or background K giant contaminants. The vertical dashed line 
denotes our suggested limit ($P_{mem}>80\%$) for identifying 
high-confidence cluster members.}
\end{figure}

\begin{figure*}
\epsscale{1.00}
\plotone{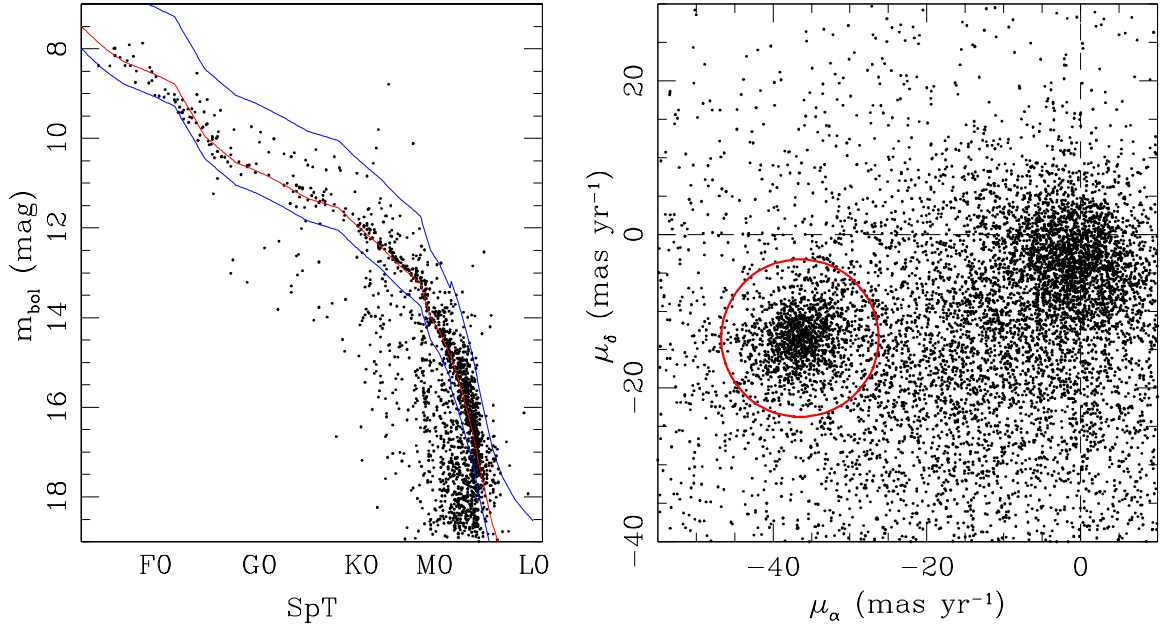}
\caption{Left: An HR diagram for all objects which have proper motions 
within 8 mas yr$^{-1}$ of the mean value for Praesepe. The field main 
sequence at the distance of Praesepe is shown with a red line; the blue 
lines outline our photometric selection limits. We identified few 
candidate members of Praesepe fainter than $m_{bol}=17.5$. The possible 
sequence below and blueward of this point is not a genuine feature, but 
is instead a result of the large number of background early-mid M dwarfs 
with similar proper motions. These stars are spatially uniformly 
distributed, which also argues that they are not associated with 
the cluster. Right: A proper motion diagram for all objects which fall 
within our photometric selection limits. The red circle outlines the 
2$\sigma$ limit for a low-mass (M5) Praesepe member.}
\end{figure*}

\begin{figure*}
\epsscale{1.00}
\plotone{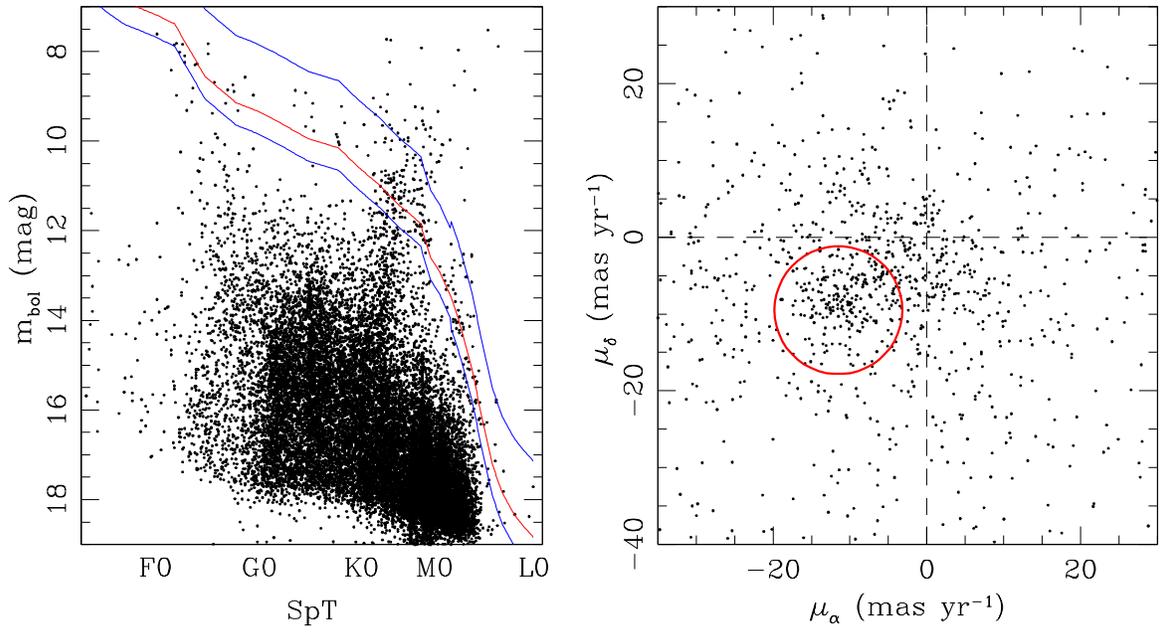}
\caption{As in Figure 4, but for Coma Ber.}
\end{figure*}

\subsection{Completeness}

As we describe in Section 2.5, there have been several previous surveys 
which identified a large number of high-confidence Praesepe members. The 
resulting sample of 381 members, comprising all stars which have been 
identified at $\ge$95\% confidence in one survey and at no lower than 
$<$80\% confidence by any others, can test the completeness of our 
proposed member list.

Of the 381 known member stars, 22 were too bright to have proper motions 
in UCAC2, so they were immediately excluded from our cluster survey. This 
suggests that most of the brightest, highest-mass stars in either cluster 
would not have been identified with our technique. Of the 359 stars which 
were not rejected due to lack of data, 330 were identified as members with 
$>$80\% confidence; the corresponding total completeness is 87\%. We found 
that 15 stars were rejected for having inconsistent photometry and 24 were 
rejected for having inconsistent proper motions. Of the 15 stars rejected 
based on their photometry, 10 also possessed discrepant proper motions, 
suggesting that these sources are probably not genuine members of Praesepe 
and raising our completeness above 90\%.

In Figure 6, we plot the completeness as a function of spectral type for 
members of Praesepe. We project that our survey is $\ga$90\% complete for 
spectral types F0 to M5, declining to 0\% completeness for spectral types 
$\le$A5 and $\ge$M7. The incompleteness for early-type stars is a result 
of the bright limit of UCAC2 data, while the incompleteness for late-type 
stars is a result of the detection limits for USNOB and 2MASS, which are 
reached nearly simultaneously for stars on the Praesepe and Coma Ber 
cluster sequences. The low-mass limit is also consistent with the results 
we summarize in Table 2 since we found no members with late M spectral 
types. We project that the 90\% completeness limits should be marginally 
later (F5 and M6) for Coma Ber since it is closer and its members are 
brighter; the completeness is also lower for K stars due to contamination 
from background K giants.

These results are mostly consistent with our comparison to individual 
surveys. In Praesepe, we find excellent agreement in comparing our list of 
high-probability candidates with those of Jones \& Stauffer (1991) and 
Hambly et al. (1995a); approximately 90\% of each survey's high-confidence 
($P_{mem}>80\%$) candidates were also identified as high-confidence 
candidates by our survey. We find less overlap with the Praesepe survey of 
Adams et al. (2002) and the Coma Ber survey of Casewell et al. (2006). Of 
the candidates which Adams et al. identify as "high-confidence" 
($P_{mem}>20\%$ and $r<4^o$), we only recovered 483 of 724 in our list of 
high-probability candidates. Casewell et al. used a moderately 
mass-dependent threshold, varying between $60\%<P_{mem}<90\%$, to identify 
60 new candidate members. Of these stars, we only recover 22.

For both of these surveys, much of the contamination can be traced to the 
use of 2MASS $JHK$ photometry in the color-selection procedures. The 
$K$,$J-K$ color-magnitude sequence for dwarfs is nearly vertical for 
spectral types M0-M6, so it is difficult to distinguish a moderately 
brighter foreground star or moderately fainter background star from a 
genuine cluster member. We found that most of the unrecovered candidates 
were background M0-M2 stars that fall below the cluster sequence in our 
HR diagrams. For the survey by Casewell et al., we also found that the 
recovery fraction was exceptionally low ($\sim$20\%) among K stars. We 
attribute this to contamination from background K giants, which affected 
both their survey and ours. We were able to identify only 13 of the 
$\sim$40 estimated K star members with high ($>$80\%) confidence (Tables 
4 and 2, respectively), suggesting that there should be only marginal 
overlap. Many of the candidates from the survey by Casewell et al. appear 
to be likely cluster members that were only identified at lower 
confidence ($50\%<P_{mem}<80\%$) by our survey. However, most of their 
remaining candidates appear to have proper motions more consistent with 
nonmovement than comovement, suggesting that they are background giants.

\begin{figure}
\epsscale{1.00}
\plotone{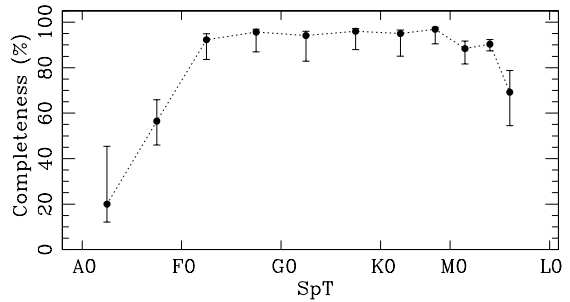}
\caption{Completeness as a function of spectral type for our 
high-confidence sample of Praesepe members. The high-mass cutoff is a 
result of image saturation, while the low-mass cutoff is a result of 
nondetection by 2MASS and USNOB. We expect similar results for Coma Ber, 
but given that its members are $\sim$1.5 magnitudes brighter, the 90\% 
completeness range will shift to later spectral types (F5-M6).} 
\end{figure}

\section{The Structure and Evolution of Praesepe and Coma Ber}

Open clusters are thought to be the birthplaces of most stars, so cluster 
evolution plays a key role in setting the environment for early stellar 
evolution. Present-day cluster properties can be used to determine their 
past history and extrapolate their future lifetime; the three most 
important sets of properties are the spatial structure (as inferred from 
mass segregation), the cluster's stellar mass function, and the total 
cluster mass.

\begin{figure}
\epsscale{1.00}
\plotone{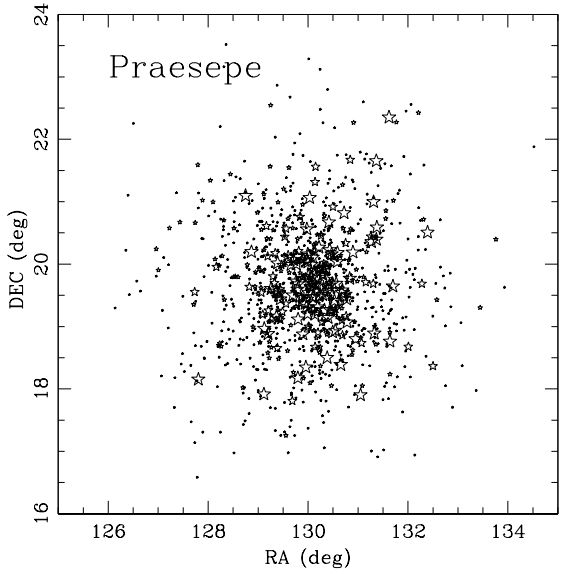}
\caption{The spatial distribution of high-probability ($P_{mem}>80\%$) 
members of Praesepe. The points are scaled to decreasing size for A-F, G, 
K, and M stars.}
\end{figure}

\begin{figure}
\epsscale{1.00}
\plotone{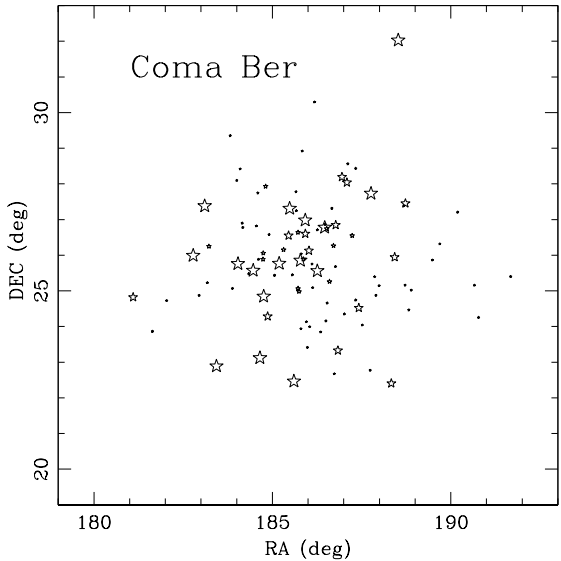}
\caption{As in Figure 6, but for Coma Ber.}
\end{figure}

\subsection{Radial Distributions and Mass Segregation}

In Figures 7 and 8, we plot the spatial distribution of all 
high-probability candidate members of Praesepe and Coma Ber. In each plot, 
we have scaled the points to decreasing sizes for A-F, G, K, and M stars. 
These figures clearly illustrate the radial density profile of each 
cluster. However, it is perilous to infer cluster properties directly from 
the distribution of individual stars. The surface density as a function of 
radius, $\Sigma(r)$, is biased in our sample because each star's radial 
distance is factored into its membership probability.

Ideally, cluster properties should be estimated using an unbiased method. 
Our parametric determination of the e-folding scale radius $r_0$ provides 
a natural diagnostic for quantifying the radial distribution and mass 
segregation of each cluster. This quantity allows us to study these 
properties without dependence on potentially biased measurements for 
individual stars, plus we can avoid arbitrary choices like the selection 
of a cutoff in $P_{Mem}$.

In Figure 9, we plot the mass-dependent function $r_0(M)$ for Praesepe 
(top) and Coma Ber (bottom). The uncertainties and upper limits were 
derived using the Monte Carlo methods described in Section 3.3. As we 
described in Section 4.2, the completeness of our sample drops for 
spectral types later than M5 in Praesepe and M6 in Coma Ber, so we do not 
plot results below these limit. In Praesepe, the scale radius increases 
significantly across the full mass range, following the power law 
$r_0$$\propto$$M^{-0.25\pm0.06}$, which indicates the clear presence of 
mass segregation. Coma Ber shows no clear trend to indicate mass 
segregation, but the result is more uncertain: 
$r_0$$\propto$$M^{-0.10\pm0.09}$. We expect Coma Ber to be less segregated 
than Praesepe due to its younger age and lower stellar density, but a 
trend with the same slope as in Praesepe is inconsistent by only 
$<$2$\sigma$.

\begin{figure}
\epsscale{1.00}
\plotone{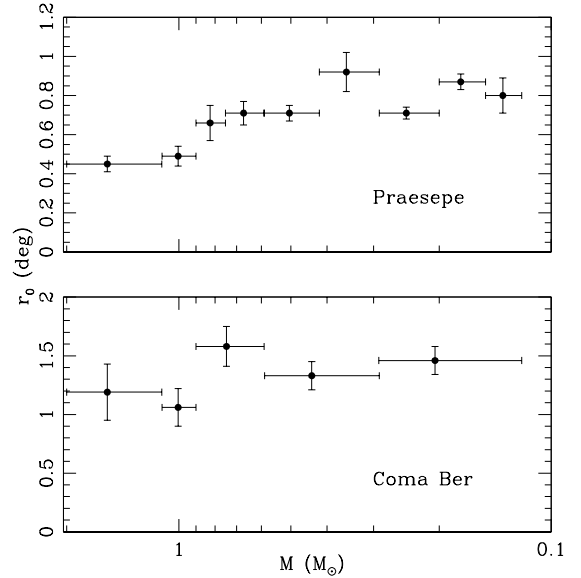}
\caption{Scale radius $r_0(M)$ for each cluster. The scale radius in 
Praesepe clearly increases with decreasing mass, indicating the presence 
of mass segregation. The corresponding trend for Coma Ber is inconclusive 
due to low number statistics.} 
\end{figure}

\subsection{Mass Functions}

The present-day mass function provides an important test of the 
evolutionary state of each cluster, assuming clusters form with a common 
initial mass function. Dynamical evolution (mass segregation and tidal 
stripping) will preferentially remove low-mass cluster members, so evolved 
clusters should show large deficits of low-mass stars. The mass function 
is defined as $\Psi(M)=dN/dM$, such that $\Psi(M)$ is the number of stars 
with masses in the interval $(m,m+dm)$. We have constructed mass functions 
using the spectral type intervals defined in Section 3.3, where the number 
of stars is the quantity $N_c$ determined in our fitting routine. These 
mass bins have uneven width, so we normalized each value to represent the 
number of stars per interval 0.1 $M_{\sun}$.

In Figure 10, we plot the cluster mass functions for Praesepe (top) and 
Coma Ber (bottom). Each function can be fit with a single power law, 
$\Psi\propto$$M^{-\alpha}$, where $\alpha=1.4\pm0.2$ for Praesepe and 
$\alpha=0.6\pm0.3$ in Coma Ber. Both power laws are significantly shallower 
than a Salpeter IMF ($\alpha=2.35$), but the Praesepe power law agrees well 
with the present-day mass function for nearby field stars 
($\alpha=1.35\pm0.2$ for 1.0-0.1 $M_{\sun}$; Reid et al. 2002).  Previous 
studies of the mass function for young clusters and unbound associations 
have also found similar slopes in this mass range 
($\alpha$$\sim$1.25$\pm$0.25; Hillenbrand 2004 and references therein).

Neither cluster has a sharp decline in the number of low-mass members 
within the mass range of our sample. Chappelle et al. (2005) found that the 
Praesepe mass function may drop sharply just below the limit of our survey 
($\la$0.12 $M_{\sun}$), which could denote the effect of tidal stripping of 
low-mass members, but we can not confirm or disprove this result. The 
shallower power law of the Coma Ber mass function suggests that some of its 
low-mass members may have been removed, but it appears that any limit for 
the total depletion of cluster members must lie below $\sim$0.12 $M_{\sun}$ 
as well.

\begin{figure}
\epsscale{1.00}
\plotone{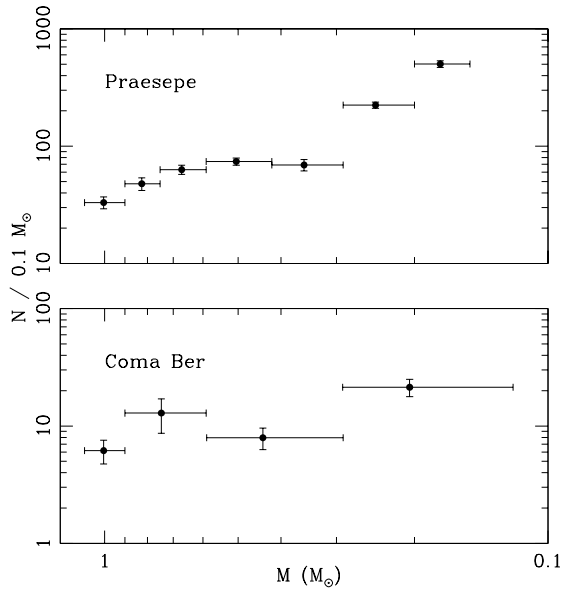}
\caption{Mass functions, $\Psi(M)=dN/dM$, for Praesepe and Coma Ber. We 
derived these results from our best-fit values for $N_c(M)$, as described 
in Section 3.3 and Table 2; each spectral type bin corresponds to a 
different width in mass, so we normalized all bins to report the number of 
stars per 0.1 $M_{\sun}$.}
\end{figure}

\subsection{Cluster Masses and Tidal Radii}

We have derived the total masses of each cluster by integrating the mass 
functions that we described in the previous section. Since these mass 
functions do not include high-mass stars, we have manually added the 
masses of known high-mass cluster members which were not identified in 
our survey, comprising $\sim$1/3 of the total mass. We identified the 
missing Praesepe members using our high-confidence cluster sample 
(Section 2.5), plus the five evolved giant members identified by 
Klein-Wassink (1927), while the corresponding members of Coma Ber were 
identified from the original member list of Trumpler et al. (1937).

We have not included any of the candidate Coma Ber members suggested by 
subsequent surveys (Bounatiro 1993; Odenkirchen et al. 1998) since it has 
been suggested that a significant fraction of these candidates may be 
spurious (Ford et al. 2001). We also did not attempt to include any 
substellar or near-substellar members of Praesepe or Coma Ber since they 
are not thought to comprise a significant fraction of the cluster mass 
(e.g. Chappelle et al. 2005).

Based on this analysis, we estimate that the total stellar populations for 
Praesepe and Coma Ber consist of 1050$\pm$30 stars earlier than M5 and 
145$\pm$15 stars earlier than M6, respectively. The corresponding total 
masses are 550$\pm$40 $M_{\sun}$ and 112$\pm$16 $M_{\sun}$. Given these 
cluster masses, we can also estimate the tidal radius of each cluster:

\[r_{t}=[\frac{GM_c}{4A(A-B)}]^{1/3}\]

(King 1962), where A and B are the Oort constants ($A=14.4$ km s$^{-1}$ 
kpc$^{-1}$; $B=-12.0$ km s$^{-1}$ kpc$^{-1}$; Kerr \& Lynden-Bell 1986). We 
derive estimated tidal radii of 11.5$\pm$0.3 pc (3.5$\pm$0.1$^o$) for Praesepe 
and 6.8$\pm$0.3 pc (4.3$\pm$0.2$^o$) for Coma Ber. In both cases, these 
radii are approximately half the radius of our search area (7$^o$). This 
suggests that our survey should be spatially complete for all bound 
members.

Finally, we note that all of these results are likely to be marginally 
underestimated due to unresolved stellar multiplicity. Given the typical 
binary frequency found for open clusters ($\sim$30\%; Patience et al. 2002) 
and the mean mass ratio for binaries ($\sim$0.3-0.7), the magnitude of this 
mass underestimate should be $\sim$20\%. We will address this problem in a 
future publication that specifically studies stellar multiplicity in both 
clusters.

\section{Summary}

We have combined archival survey data from the SDSS, 2MASS, USNOB1.0, and 
UCAC-2.0 surveys to calculate proper motions and photometry for $\sim$5 
million sources in the fields of the open clusters Praesepe and Coma Ber. 
Of these sources, 1010 stars in Praesepe and 98 stars in Coma Ber have 
been identified as candidate members with probability $>$80\%; 442 and 61, 
respectively, are newly identified as high-probability candidates for the 
first time. We estimate that this survey is $>$90\% complete across a wide 
range of spectral types (F0 to M5 in Praesepe, F5 to M6 in Coma Ber).

We have also investigated each cluster's mass function and the stellar 
mass dependence of their radii in order to quantify the role of mass 
segregation and tidal stripping in shaping the present-day mass function 
and spatial distribution. Praesepe shows clear evidence of mass 
segregation, but if significant tidal stripping has occurred, it has 
affected only members near and below the substellar boundary ($\la$0.15 
$M_{\sun}$). Low number statistics make it difficult to quantify the level 
of mass segregation in Coma Ber. The shallower slope of its mass function 
suggests that some mass loss has occurred, but any mass limit for total 
depletion of the cluster population must fall below the limit of our 
survey.

\acknowledgements

The authors thank John Stauffer for providing helpful feedback on the 
manuscript. This work makes use of data products from 2MASS, a joint 
project of the University of Massachusetts and IPAC/Caltech, funded by 
NASA and the NSF. Our research has also made use of the USNOFS Image and 
Catalogue Archive operated by the USNO, Flagstaff Station 
(http://www.nofs.navy.mil/data/fchpix/). Funding for the SDSS has been 
provided by the Alfred P. Sloan Foundation, the Participating 
Institutions, the NSF , the U.S. DoE, NASA, the Japanese Monbukagakusho, 
and the Max Planck Society, and the Higher Education Funding Council for 
England. The SDSS is managed by the Astrophysical Research Consortium for 
the Participating Institutions.

\appendix

\section{Stellar SED Library}

There is no single source in the literature that describes all of the SED 
data that we require, so we compiled a preliminary set of models from a 
heterogeneous set of empirical observations. We then optimized these 
models by comparing the color-magnitude sequences to the single-star 
sequence of our high-confidence Praesepe sample (Section 2.5).

Luminosities and optical colors for our high-mass and intermediate-mass 
stellar models (spectral types B8 to K7) were based on the absolute UBV 
magnitudes of Schmidt-Kaler (1982), which we converted to SDSS absolute 
magnitudes using the color transformations of Jester et al. (2005). We 
then used the optical-NIR colors ($V-K$, $J-K$, and $H-K$) of Bessell and 
Brett (1988) to estimate $JHK$ absolute magnitudes, and converted these 
values to the 2MASS filter system using the NIR color transformations of 
Carpenter et al. (2001). We estimated absolute bolometric magnitudes for 
each model using the bolometric corrections of Masana et al. (2005).

For M dwarfs (M0-L0), we based our models on the fourth-order polynomial 
relation of absolute $JHK$ vs spectral type described by Cruz et al. 
(2007); they only explicitly defined this relation for spectral types 
later than M6, so we used 2MASS observations of stars in the CNS3 catalog 
(Gliese \& Jahreiss 1991) and the 8 pc sample (Reid et al. 2002) to 
estimate the appropriate polynomial relation for M0-M5 stars. We combined 
these results with the $r-i$, $i-z$, and $z-J$ colors of West et al. 
(2005) and the $u-g$ and $g-r$ colors of Bochanski et al. (2007). We 
estimated absolute bolometric magnitudes using the bolometric corrections 
of Leggett (1992) and Leggett et al. (2002).

Finally, we optimized our set of spectral type models by comparing 
theoretical color-color and color-magnitude sequences to the empirical 
color-color and color-magnitude sequences of our sample of high-confidence 
Praesepe members. We found that the absolute magnitudes of our models 
differed from the empirical sequence at spectral types F2-F8 and at the 
K/M boundary, so we adjusted these absolute magnitudes to match the 
empirical sequences. We did not find any need to adjust the colors of any 
model, which suggests that any discrepancies are a result of the 
bolometric corrections.

In Table 5, we list our final set of spectral type models. Our fitting 
routine subsamples this model grid by linearly interpolating to predict 
values for intermediate spectral types; our final grid of models (491 in 
all) proceeds from B8 to L0 in steps of 0.1 subclasses, following the 
recent nomenclature trend to proceed directly from K5 to K7 to M0, not 
using subclasses K6, K8, or K9.

For high-mass stars ($\le$F2), we directly adopted masses from the models 
of Schmidt-Kaler (1982). For lower-mass stars, we adopted effective 
temperatures for each model using the dwarf temperature scales of 
Schmidt-Kaler (1982) (for spectral types $\le$M0) and Luhman (1999) (for 
spectral types $>$M0). We then combined these $T_{eff}$ values with the 
500 Myr isochrones of Baraffe et al. (1998) to estimate stellar masses. 
The appropriate mixing length has been found to change with mass (Yildiz 
et al. 2006), so for masses $>$0.6 $M_{\sun}$, we used the models with a 
mixing length of $H_P$. For masses $<$0.6 $M_{\sun}$, we used the models 
with a mixing length of 1.9 $H_P$. 

Several studies (e.g. Hillenbrand \& White 2004; Lopez-Morales 2007) have 
found that theoretical models can underpredict masses, so these values 
should be considered with some caution. The most uncertain mass range is 
$<$0.5 $M_{\sun}$. Observational calibrations suggest that the models 
underpredict masses by $\sim$10-20\% in the mass range of 0.2-0.5 
$M_{\sun}$, and the models are almost completely uncalibrated for lower 
masses. We have addressed this problem by increases the masses of M1 stars 
by 5\%, M2 stars by 10\%, and later-type stars by 20\%; these adopted 
values are more consistent with the observations (e.g. Lacy 1977; Delfosse 
et al. 1999; Creevy et al. 2005; Lopez-Morales \& Ribas 2005).

We list all of the adopted values of $M$ and $T_{eff}$ in Table 5.

\begin{deluxetable*}{lrrrrrrrrrrl}
\tabletypesize{\scriptsize}
\tablewidth{0pt}
\tablecaption{Stellar SEDs}
\tablehead{\colhead{SpT} & \colhead{$M_u$} & \colhead{$M_g$} & 
\colhead{$M_r$} & \colhead{$M_i$} & \colhead{$M_z$} & 
\colhead{$M_J$} & \colhead{$M_H$} & \colhead{$M_K$} & 
\colhead{$M_{bol}$} & \colhead{$T_{eff}$} & \colhead{$M$ ($M_{\sun}$)}
}
\startdata
B8&0.32&-0.39&-0.04&0.34&0.62&0.01&0.10&0.11&-1.00&11900&3.8\\
A0&1.58&0.47&0.72&1.04&1.28&0.54&0.58&0.56&0.30&9520&2.9\\
A2&2.41&1.22&1.39&1.65&1.87&1.12&1.15&1.12&1.10&8970&2.4\\
A5&3.14&1.88&1.95&2.15&2.32&1.53&1.52&1.48&1.75&8200&2.0\\
A7&3.47&2.21&2.23&2.40&2.55&1.75&1.71&1.66&2.08&7580&1.8\\
F0&3.94&2.77&2.68&2.79&2.90&2.10&2.01&1.96&2.61&7200&1.6\\
F2&4.23&3.10&2.96&3.04&3.13&2.32&2.20&2.14&2.89&6890&1.5\\
F5&5.01&3.90&3.68&3.69&3.74&2.85&2.67&2.61&3.61&6440&1.25\\
\enddata
\end{deluxetable*}

\clearpage
\LongTables

\begin{deluxetable*}{lrrrrrrl}
\tablenum{3}
\tabletypesize{\scriptsize}
\tablewidth{0pt}
\tablecaption{Candidate Members of Praesepe}
\tablehead{\colhead{ID} & \colhead{SpT} & \colhead{$m_{bol}$} &
\colhead{$\mu_{\alpha}$} & \colhead{$\mu_{\delta}$} & 
\colhead{$\sigma_{\mu}$} & \colhead{$P_{mem}$} & \colhead{Previous ID}
\\
\colhead{} & \colhead{} & \colhead{(mag)} & 
\multicolumn{3}{c}{(mas yr$^{-1}$)} & \colhead{(\%)}
}
\startdata
2MASS J08401535+1959394&A4.6$\pm$4.9&8.37$\pm$0.02&-35.8&-12.3&0.7&100.0&KW271 (F1; Abt 1986)\\
2MASS J08405693+1956055&A5.7$\pm$2.1&8.38$\pm$0.02&-36.1&-15.4&1.2&100.0&KW350 (A8; Abt 1986)\\
2MASS J08462889+2221079&A6.1$\pm$2.3&8.00$\pm$0.02&-38.3&-14.8&0.5&88.7&\\
2MASS J08383786+1959231&A6.2$\pm$2.0&7.99$\pm$0.01&-37.4&-13.6&0.7&100.0&KW114 (A8; Abt 1986)\\
2MASS J08390359+1959591&A6.2$\pm$2.0&8.15$\pm$0.01&-34.2&-13.3&0.6&100.0&KW143 (A8; Abt 1986)\\
2MASS J08421080+1856037&A6.3$\pm$1.9&7.74$\pm$0.01&-34.1&-12.1&1.1&99.2&KW449 (A7; Abt 1986)\\
2MASS J08411377+1955191&A6.7$\pm$1.9&8.19$\pm$0.01&-36.9&-12.6&0.6&100.0&KW375 (A7; Abt 1986)\\
2MASS J08364800+1852580&A7.2$\pm$2.3&8.38$\pm$0.01&-35.4&-13.6&0.9&99.8&KW538 (A9; Abt 1986)\\
\enddata
\tablecomments{The full version of Table 3 will be published as an online-only table in ApJ.}
\end{deluxetable*}

\clearpage

\begin{deluxetable*}{lrrrrrrl}
\tablenum{4}
\tabletypesize{\scriptsize}
\tablewidth{0pt}
\tablecaption{Candidate Members of Coma Ber}
\tablehead{\colhead{ID} & \colhead{SpT} & \colhead{$m_{bol}$} &
\colhead{$\mu_{\alpha}$} & \colhead{$\mu_{\delta}$} & 
\colhead{$\sigma_{\mu}$} & \colhead{$P_{mem}$} & \colhead{Previous ID}
\\
\colhead{} & \colhead{} & \colhead{(mag)} & 
\multicolumn{3}{c}{(mas yr$^{-1}$)} & \colhead{(\%)}
}
\startdata
2MASS J12175090+2534167&F0.3$\pm$2.0&7.61$\pm$0.01&-10.8&-10.0&0.6&100.0&Tr 49 (F3; Abt \& Levato 1977)\\
2MASS J12122488+2722482&F2.0$\pm$1.7&7.85$\pm$0.01&-12.2&-9.1&0.8&99.9&Tr 19 (F5; Abt \& Levato 1977)\\
2MASS J12160837+2545373&F2.2$\pm$1.9&7.82$\pm$0.02&-10.9&-9.5&0.7&99.9&Tr 36 (F3; Abt \& Levato 1977)\\
2MASS J12255195+2646359&F2.5$\pm$2.0&8.02$\pm$0.02&-12.9&-8.8&0.7&100.0&Tr 118 (F6; Abt \& Levato 1977)\\
2MASS J12134391+2253168&F2.8$\pm$1.9&7.84$\pm$0.02&-12.6&-9.4&0.6&99.6&Bou 38 (F5V; Bounatiro 1993)\\
2MASS J12234101+2658478&F3.0$\pm$1.8&8.10$\pm$0.02&-11.7&-8.1&0.6&100.0&Tr 101 (F6; Abt \& Levato 1977)\\
2MASS J12310309+2743491&F3.0$\pm$1.7&8.25$\pm$0.02&-12.7&-7.6&0.7&99.6&Tr 162 (F7; Abt \& Levato 1977)\\
2MASS J12222475+2227509&F4.0$\pm$1.5&8.30$\pm$0.01&-11.6&-9.3&0.6&99.8&Tr 90 (F6; Abt \& Levato 1977)\\
\enddata
\tablecomments{The full version of Table 4 will be published as an online-only table in ApJ.}
\end{deluxetable*}

\end{document}